\documentclass[twocolumn,prl,aps]{revtex4-2}
\usepackage{floatflt}
\usepackage{psfrag}
\usepackage{amsmath}   
\usepackage{amssymb}
\usepackage{amsfonts}
\usepackage{mathrsfs}
\usepackage{graphicx}

\usepackage[dvipsnames]{xcolor}
\usepackage[colorlinks=true, citecolor=Cerulean,linkcolor=blue, urlcolor = Cerulean]{hyperref}

\begin{document}

\title
{Hybridization of Ferromagnetic and Cyclotron Resonances in a Two-Dimensional Electron System on a Ferromagnetic Film}

\author{A. A. Zabolotnykh$^{a,b}$, I. V. Zagorodnev$^{a,b}$, A. A. Matveev$^{a,b}$, D. A. Rodionov$^{a}$, O. Yu. Arkhipova$^{a,b,c}$, D. V. Kalyabin$^{a,b}$, A. R. Safin$^{a,b,d}$, S. A. Nikitov$^{a,b,e}$}
\affiliation{$^a$ Kotelnikov Institute of Radio-engineering and Electronics of the RAS, Mokhovaya 11-7, Moscow 125009 Russia\\
$^b$ Moscow Institute of Physics and Technology (National Research University), Dolgoprudny, Moscow region, 141701 Russia \\
$^c$ Bauman Moscow State Technical University, Moscow, 105005 Russia\\
$^d$ National Research University “Moscow Power Engineering Institute,” Moscow, 111250 Russia \\
$^e$ Laboratory of Magnetic Metamaterials, Saratov State University, Saratov, 410012 Russia}

\date{January 15, 2026}

\begin{abstract}
The microwave response of a two-dimensional (2D) electron system located on a dielectric ferromagnetic film, which in turn lies on a conducting metal (gate), has been theoretically studied. The entire system has been placed in the perpendicular static magnetic field. It has been found that the ferromagnetic resonance of the film and the cyclotron resonance of the electrons of the 2D system interact in the magnetic field, leading to their repulsion (“anticrossing”). It has been revealed that the anticrossing region is characterized not only by the modification of resonance frequencies compared to the cyclotron resonance in the 2D system without the ferromagnetic substrate and the ferromagnetic resonance in the film without the 2D system, but also by a strong change in the resonance linewidths.
\end{abstract}

\maketitle

Spintronic devices based on ferromagnetic materials can be used to solve various applied problems~\cite{Best_Review, Shao, Chappert}. For example, thin ferromagnetic films are proposed for magnetic field detection \cite{Filimonov2022} and neuromorphic computing \cite{Sadovnikov_2018}, as inverters, phase shifters, and magnonic logic elements \cite{Wave_base, Roadmap_SW}. Dielectric ferromagnets with low intrinsic losses, such as yttrium iron garnet (YIG), are often used in such devices \cite{Onbasli2014, Best_Review}. Of particular scientific interest are samples of iron garnets possessing the perpendicular magnetic anisotropy (PMA) field, allowing the induction of stable out-of-plane orientation of magnetic moments, which can be used to develop memory elements and logic devices \cite{Xu2022, Das2024, Zhang2022, Wave_base}.

Modern scientific investigations are focused on the study of the resonance properties of not only single ferromagnet, but also heterostructures \cite{Best_Review, Heins2025, Shao}. In particular, ferromagnet/heavy metal structures are proposed for the detection of microwaves of various polarizations \cite{Luig_circ, Emori2018}. Coupling between the magnetic and elastic subsystems in a YIG --- piezoelectric heterostructure makes it possible to solve problems of multiplexing microwave signals \cite{Sadovnikov_2017}. About 30 years ago, ferromagnet --– two-dimensional (2D) electron system structures (GaAs quantum wells) were considered and manufactured, in which the possibility of reading the state of the ferromagnet by measuring the Hall resistance of the 2D system was demonstrated \cite{Monzon1997, Meier2000}. In recent years, the possible hybridization of plasma and ferro- or antiferromagnetic resonance of such systems has been actively discussed \cite{bludov2019,pikalov2021,costa2023, Yuan2024,kuznetsov2025,Yuan2025}. However, 2D systems also has another fundamental and practically important excitation called cyclotron resonance \cite{Chiu:1976,Theis1977,Ando1982}. A distinctive feature of 2D electron systems is the ability to control the carrier density using a nearby metal electrode (gate). The ability to vary the carrier density by applying a gate voltage allows for fine tuning of cyclotron resonance conditions, as recently demonstrated (theoretically and experimentally) in \cite{Zabolotnykh:2021CR,Muravev:2024CR}. To the best of our knowledge, resonance properties associated with the hybridization of cyclotron and ferromagnetic resonances have not yet been discussed, although they can be used to develop new hybrid elements for spintronics and microwave photonics. Thus, the properties of both the ferromagnetic and cyclotron resonances in a hybrid system will critically depend on the parameters of the 2D system. For instance, a sufficiently high-quality ferromagnetic resonance can increase the quality factor of cyclotron resonance (as will be demonstrated below), allowing, for example, its observation at higher temperatures or in “dirty” systems. Furthermore, since both resonances are sensitive to the magnetic field, its variation makes it possible to change the coupling strength between the magnetic and electronic subsystems.

In the present paper, we study the manifestation of the considered resonances in the response to a circularly polarized electromagnetic wave incident normally on a structure consisting of 2D electron system, a dielectric ferromagnetic film, and a perfectly conducting metal.

\begin{figure}
    \centering
\includegraphics[width=\linewidth]{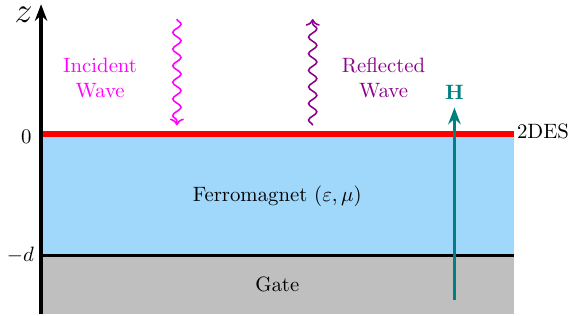}
    \caption{Scheme of the structure under study.
    }
    \label{Fig:sys}
\end{figure}

We consider the 2D system located in the $z = 0$ plane. The ferromagnetic film with the frequency-independent permittivity $\varepsilon$ and the permeability tensor $\hat \mu=\hat \mu(\omega)$ whose frequency dependence is discussed later, see Eq.~(\ref{Eq:mu}), occupies the region $0 > z > -d$ while the perfectly conducting metal gate is located in the region $z< -d$, see Fig.~\ref{Fig:sys}. The circularly polarized electromagnetic wave $E_0 e^{-i\omega z /c -i\omega t}$ falls normally on the system from the vacuum region $z > 0$. The reflected wave has the form $E_r e^{i\omega z /c -i\omega t}$ and is also circularly polarized. Henceforth, we assume that $\omega>0$. Our goal is to determine the frequency dependence of the amplitude reflection coefficient $r=E_r/E_0$, as well as the energy absorption coefficient, which is equal to $A=1-|r|^2$, because a wave transmitted into the ideal metal gate is absent.

To determine the reflection coefficient, the electric fields of the waves in the regions $z > 0$ and $0 > z > -d$ are matched using boundary conditions at $z = 0$ and $z=-d$. As discussed above, the wave at $z > 0$ has the form $E_0 e^{-i\omega z /c -i\omega t}+E_r e^{i\omega z /c -i\omega t}$. At $0 > z > -d$, we have $E_1 e^{-i\omega z \sqrt{ \varepsilon \mu_{\pm}} /c -i\omega t}+E_2 e^{i\omega z \sqrt{\varepsilon \mu_{\pm}} /c -i\omega t}$, where $E_{1,2}$ are the wave amplitudes in the film, while $\mu_{\pm}$ are the permeabilities that correspond to circular polarizations and are discussed below.

Boundary conditions are as follows. The first condition includes the continuity of the tangential field component at the $z = 0$ interface and its vanishing at the surface of the perfectly conducting gate $E(z=-d)=0$. The second condition relates the jump of the tangential magnetic field component at the $z = 0$ interface, which is caused by the presence of currents in the 2D system and can be expressed in terms of the jump of the derivative of the electric field at this interface, to the current density $j^{2D}$ in the 2D system (the CGS system of units is used throughout the paper):
\begin{equation}
    \frac{1}{\mu_{\pm}(z)}\partial_z E(z)|^{+0}_{-0}=-\frac{4 \pi i\omega}{c^2}j^{2D}.
\end{equation}
In fact, it is the latter equation that couples the ferromagnetic and two-dimensional subsystems. The current density is related to the electric field via local Ohm’s law: $j_{\pm}^{2D}=\sigma_{\pm}E_{\pm}(z=0)$, where $\sigma_{\pm}$ are the 2D circular conductivities.

Substituting the solutions in the film and vacuum into the boundary conditions, we find the amplitude reflection coefficient:
\begin{equation}
\label{Eq:refl}
    r_\pm =\frac{E_r}{E_0}= \frac{1-i\sqrt{\frac{\varepsilon}{\mu_\pm}}\cot \left(\frac{\omega d}{c}\sqrt{\varepsilon\mu_\pm}\right) - \frac{4\pi\sigma_\pm}{c}}{1+i\sqrt{\frac{\varepsilon}{\mu_\pm}}\cot \left(\frac{\omega d}{c}\sqrt{\varepsilon\mu_\pm}\right) + \frac{4\pi\sigma_\pm}{c}}.
\end{equation}

To determine the frequency dependence of the amplitude reflection coefficient $r_{\pm}$ and the energy absorption coefficient $A_{\pm}$, it is necessary to specify the explicit form of the 2D conductivity $\sigma_{\pm}$ and the permeability $\mu_{\pm}$ of the film. The permeability can be found using the following relationship:
\begin{equation}
\label{Eq:mu_chi}
    \mu_\pm = \mu_{xx} \mp i\mu_{xy} = 1 + 4\pi\left(\chi_{xx} \mp i\chi_{xy}\right).
\end{equation}
Here, $\chi_{xx}$ and $\chi_{xy}$ are the components of the magnetic susceptibility tensor, which are presented in the explicit form in Refs.~\cite{Best_book, Vanderveken2022}. Let the sample under consideration have the PMA field and the ground state of magnetization coincide with the direction of the applied static magnetic field $\mathbf{H}$. In this case, we obtain
\begin{equation}
\label{Eq:mu}
    \mu_\pm = 1+\frac{\omega_M}{\omega_1\pm \omega -i\alpha\omega}, 
\end{equation}
where $\omega_M = \gamma \cdot 4\pi M_s$, $\gamma = g e / \left(2m_0c\right)$ is the gyromagnetic ratio, $g$ is the Landé factor, $m_0$ is the mass of a free electron, and $M_s$ is the saturation magnetization; $\omega_1 = \gamma \left( H+H_p-4\pi M_s \right)$, where $H_p$ is the magnitude of the PMA field; and $\alpha$ is the Gilbert damping coefficient~\cite{Onbasli2014}.

According to the dynamic (optical) Drude model, the conductivity of the 2D system is given by the formula
\begin{equation}
\label{Eq:cond}
    \sigma_\pm =  \frac{iD}{\omega \pm  \omega_c+i/\tau},
\end{equation} 
Here, $D=ne^2/m^*$ is the Drude weight, where $n$, $m^*$, and $\tau$ are respectively the density, the effective mass, and the relaxation time of electrons in the 2D system, $e$ is the elementary charge, $\omega_c =eH/m^*c$ is the cyclotron frequency; note that the effective mass, for example, in graphene, may depend on the Fermi energy.

It is important to note that ferromagnetic and cyclotron resonances occur for the same circular polarization of the incident radiation. Indeed, Eqs.~(\ref{Eq:mu}) and (\ref{Eq:cond}) show that resonances in our notation occur for the “$-$” polarization and are absent for the “$+$” polarization.

Further, we assume for simplicity that the characteristic frequencies and, accordingly, the magnetic fields are small compared to the characteristic frequency of the Fabry–Perot cavity created by the film:
\begin{equation}
    \omega d  \sqrt{\varepsilon \mu_{\pm}}/c \ll 1.
\end{equation}
Under this condition, we can use the approximation
\begin{equation}
\label{Eq:appr_cot}
    \cot \left(\frac{\omega d}{c}\sqrt{\varepsilon\mu_\pm}\right) \approx \frac{c}{\omega d\sqrt{\varepsilon\mu_\pm}}
\end{equation}
which simplifies the expression for $r_{\pm}$~(\ref{Eq:refl}). Note that after the use of this approximation, Eq.~(\ref{Eq:refl}) does not contain the permittivity of the film $\varepsilon$, which emphasizes the magnetic nature of the considered phenomena in this limit.

Substituting Eqs. (\ref{Eq:mu}) and (\ref{Eq:cond}) into Eq. (\ref{Eq:refl}) for the reflection coefficient and using approximation (\ref{Eq:appr_cot}), we can obtain the absorption coefficient $A=1-|r_-|^2$ for the wave with the “$-$” polarization in the explicit form
\begin{equation}
\label{Eq:Absorb}
    A=\frac{4R}{(1+R)^2+I^2}.
\end{equation}
Here, $R$ is the coefficient qualitatively responsible for the absorption of the wave by the system and is given by the expression
\begin{equation}
\label{Eq:R}
    R=\frac{\alpha \omega_M c/d}{(\omega_1+\omega_M-\omega)^2+\alpha^2\omega^2}+\frac{\Gamma/\tau}{(\omega-\omega_c)^2+\tau^{-2}},
\end{equation}
while $I$ is the coefficient determining the position of
the resonance and has the form
\begin{multline}
\label{Eq:I}
    I=\frac{c}{\omega d}\left(1-\frac{\omega_M(\omega_1+\omega_M-\omega)}{(\omega_1+\omega_M-\omega)^2+\alpha^2\omega^2} \right)+ \\+\frac{\Gamma(\omega-\omega_c)}{(\omega-\omega_c)^2+\tau^{-2}}, 
\end{multline}
where the parameter $\Gamma=4 \pi e^2 n /cm^*$ has the dimension of frequency and is determined by the effective mass and density of electrons in the 2D system; note that the 2D density can be controlled by applying a voltage to the gate.

Let us qualitatively analyze Eq.~(\ref{Eq:Absorb}) for the absorption coefficient. The coefficient $R$ in the numerator of Eq.(\ref{Eq:Absorb}) is the sum of terms proportional to $\alpha$ and $1/\tau$; i.e., $R$ is responsible for the losses in the film and the 2D electron system. In addition, it is included in the denominator of Eq. (\ref{Eq:Absorb}) and, therefore, reflects the contribution to the linewidth from both the Gilbert damping in the ferromagnet and the collisional losses in the 2D electron system. Note that radiative losses lead to the appearance of unity in the denominator of Eq. (\ref{Eq:Absorb}). In the limiting case $1/\tau \to 0$ and $\alpha \to 0$, the condition $I=0$ determines the resonant frequencies, which are given by the explicit expressions as follows:
\begin{multline}
\label{Eq:res_freq}
    \omega_{r1,r2}=\frac{\omega_c+(1+\Gamma d/c)\omega_1+\omega_M \Gamma d/c}{2(1+\Gamma d/c)}\pm 
    \\ \pm \frac{\sqrt{\left((1+\frac{\Gamma d}{c})\omega_1-\omega_c +\omega_M \frac{\Gamma d}{c}\right)^2+4\omega_c \omega_M\frac{\Gamma d}{c}}}{2(1+\Gamma d/c)}.
\end{multline}
In the absence of a ferromagnet, the permeability of the film is equal to 1 (as $\omega_M=0$). In the absence of a conducting 2D system, there are no charge carriers and, accordingly, $\Gamma=0$. Therefore, the last term in the radicand proportional to $\omega_M \Gamma$ in the resulting expression is responsible for the interplay of resonances. Neglecting terms $\propto \omega_M \Gamma$, we obtain the “unperturbed” frequencies of the ferromagnetic and cyclotron resonances
\begin{equation}
\label{Eq:res0}
    \omega_{r1}^0=\omega_1, \quad \omega_{r2}^0=\omega_c/(1+\Gamma d/c),
\end{equation}
where $\omega_1$ is defined below Eq.~(\ref{Eq:mu}) as $\omega_1=\gamma \left( H+H_p-4\pi M_s \right)$. For a strong interplay of resonances to arise, their unperturbed frequencies given by Eqs.~(\ref{Eq:res0}) should be close to each other, while for anticrossing to take place, they should intersect. Let us discuss the conditions for the coincidence of the frequencies given by Eqs. (\ref{Eq:res0}) under the variation of the magnetic field $H$. The frequencies $\omega_1$ and $\omega_c$ vary as $\omega_1 \propto H ge/2m_0c$ and $\omega_c\propto H e/m^*c$. The Landé factor for ferromagnetic films usually does not differ by orders of magnitude from the value $g=2$ for YIG \cite{Onbasli2014}. For example, the Landé factor for bismuth-doped thulium garnet films with thicknesses of about several microns demonstrating a high PMA field is $g\approx 1.7$ \cite{Zhang2022}. At the same time, the relation $m^* \ll m_0$ is typical for electrons in semiconductor structures. Therefore, the ferromagnetic resonance frequency increases with the magnetic field more slowly than the cyclotron frequency. The frequencies of these resonances (in a nonzero magnetic field) can be equal, for example, for ferromagnetic films with a sufficiently high PMA field $H_p>4\pi M_s$. Then, the value of the magnetic field at the intersection point is determined as
\begin{equation}
\label{Eq:Hr}
    H_r=\frac{H_p-4\pi M_s}{e/\left( m^*c \gamma (1+\Gamma d/c)\right)-1}.
\end{equation}

\begin{figure}
    \centering
\includegraphics[width=\linewidth]{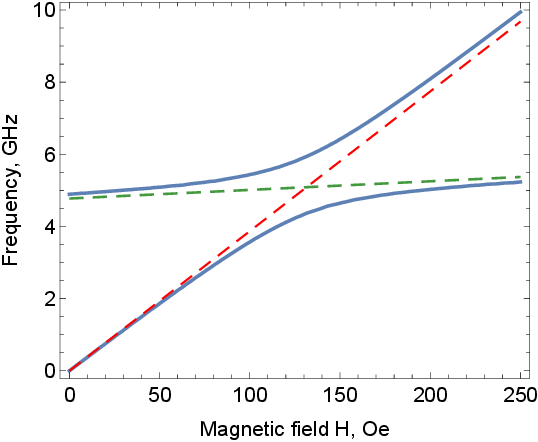}
    \caption{Resonance frequencies given by Eq.~(\ref{Eq:res_freq}) disregarding damping for the bismuth-doped thulium garnet (Bi:TmIG) film with the parameters $d=20$\, $\mu$m, $\omega_M/2\pi= 1.7$ GHz, $H_p=2700$ Oe, which corresponds to $\omega_1(H=0)/2\pi=4.8$~GHz; the typical parameters for 2D electron system based on AlGaAs quantum well were taken as follows: $m^*=0.067m_0$, $n=7.5 \cdot 10^{11}$ cm$^{-2}$, which corresponds to $\Gamma/2\pi \approx 190$~GHz and $\Gamma d /c \approx 0.08$. The red and green dashed lines indicate the frequencies of noninteracting cyclotron and ferromagnetic resonances~(\ref{Eq:res0}), respectively.
    }
    \label{Fig:freq}
\end{figure}

\begin{figure}
    \centering
\includegraphics[width=\linewidth]{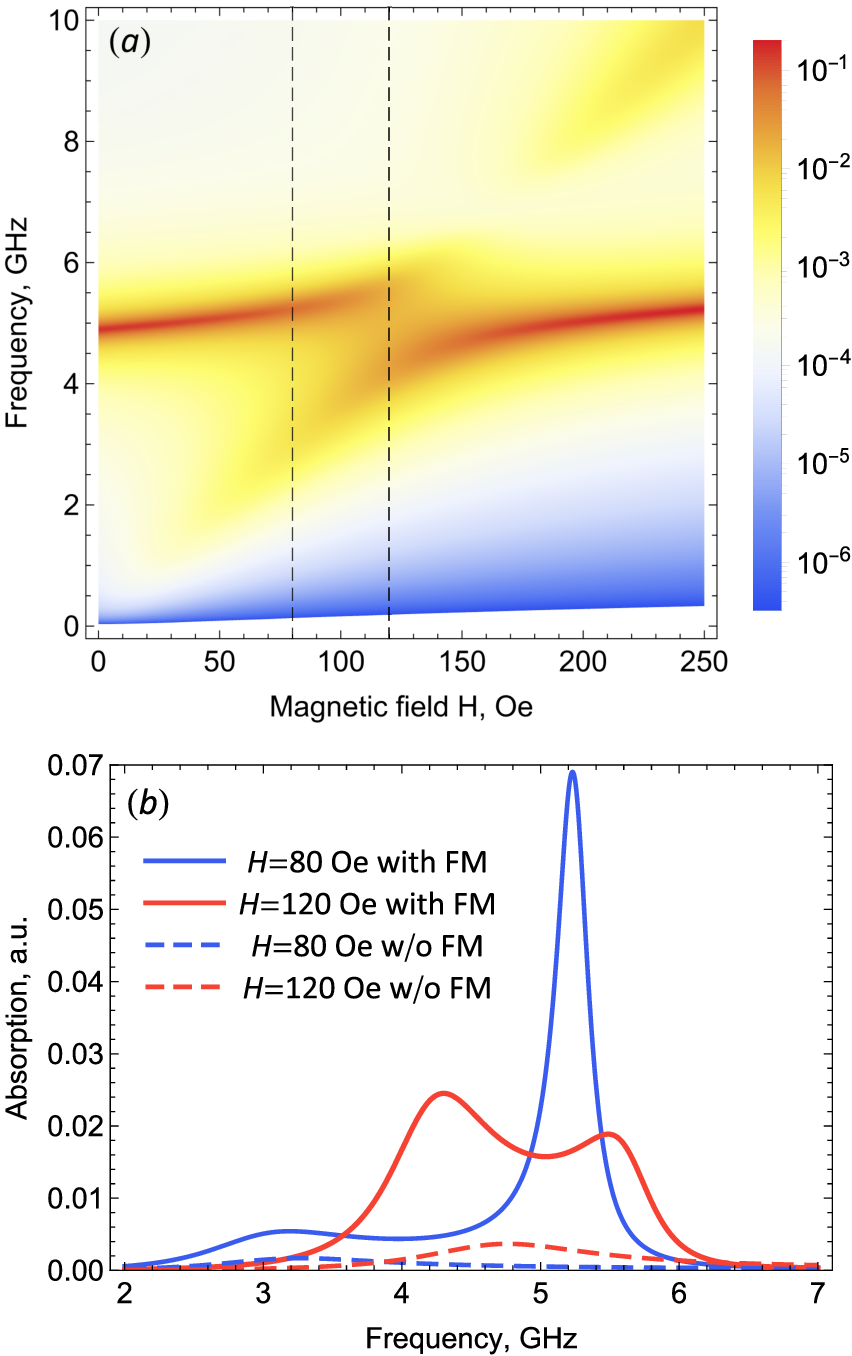}
   \caption{(a) (Magnetic field, frequency) map
for the absorption of an electromagnetic wave in the system given by Eqs.~(\ref{Eq:Absorb})--(\ref{Eq:I}). The vertical dashed lines indicate the magnetic fields for which lines in panel (b) are plotted. (b) Frequency dependencies of the absorption coefficient (solid lines) in the structure under consideration and (dashed lines) in the 2D electron system placed on the non-ferromagnetic film ($\mu_{\pm}=1$ instead of Eq.~(\ref{Eq:mu})) at magnetic fields of (blue lines) 80 and (red lines) 120 Oe. The damping coefficient of the film is $\alpha=10^{-2}$, the electron relaxation time in the 2D system is $\tau=200$ ps, and the remaining parameters are the same as in the caption to Fig.~\ref{Fig:freq}.
    }
    \label{Fig:abs}
\end{figure}
We find the difference between the resonant frequencies at $H=H_r$ by substituting Eq.~(\ref{Eq:Hr}) into
Eqs. (\ref{Eq:res_freq}):
\begin{equation}
\label{Eq:delta}
    \Delta\omega=\frac{\sqrt{(\omega_M\Gamma d/c)^2+4\omega_{cr}\omega_M\Gamma d/c}}{1+\Gamma d/c},
\end{equation}
where $\omega_{cr}=eH_r/m^*c$.

The relation $\Gamma d/c \ll 1$ is typical for realistic systems, so $\Delta\omega$ (\ref{Eq:delta}) increases with the electron density in the 2D system $n\propto \Gamma$, the frequency $\omega_M\propto M_s$ (\ref{Eq:mu}), and film thickness $d$. That is why, relatively large but realistic parameters are taken in the estimate below for the emerging anticrossing to be sufficiently pronounced. Namely, we consider the bismuth-doped thulium garnet (Bi:TmIG) film with the parameters $\gamma = 15 \cdot 10^{6} \, \text{rad}/(\text{s}\cdot\text{Oe})$, $4\pi M_s = 700$ G, and $\alpha = 1\cdot10^{-2}$ \cite{Xu2022, Zhang2022}; we assume that the film has the PMA field $H_p = 2700$ Oe and the thickness $d=20$ $\mu$m. As the 2D system, we consider a AlGaAs quantum well with $m^*=0.067m_0$, $n=7.5 \cdot 10^{11}$~cm$^{-2}$, and the electron relaxation time $\tau=200$~ps ($1/2 \pi \tau=0.8$ GHz), which corresponds to a mobility of about $5 \cdot 10^6$ cm$^2$/(V s). According to Eq. (\ref{Eq:Hr}), the anticrossing for the given parameters occurs in a field of $131$ Oe, and its magnitude is equal to 1.6 GHz according to Eq. (\ref{Eq:delta}), which is illustrated in Fig.~\ref{Fig:freq}, where the magnetic field dependencies of the resonant frequencies given by Eq. (\ref{Eq:res_freq}) are shown.

Figure~\ref{Fig:abs} shows the absorption coefficient of an electromagnetic wave at various frequencies and magnetic fields. As expected, the resonant frequencies given by Eq.~(\ref{Eq:res_freq}) and shown in Fig.~\ref{Fig:freq} correspond almost everywhere to absorption maxima. The anticrossing of the cyclotron and ferromagnetic resonances is also evident at a magnetic field of approximately 130 Oe. Figure~\ref{Fig:abs}b shows the frequency dependencies of the absorption coefficient near and far from the anticrossing. The presence of a ferromagnetic film leads to the enhancement of the cyclotron resonance response. At the same time, the ferromagnetic resonance broadens.

We also note that Fig.~\ref{Fig:abs} shows a significant suppression (“antiresonance”) of absorption in the region of a magnetic field of 165 Oe and a frequency of 7 GHz, which is a manifestation of the destructive interference of cyclotron and ferromagnetic resonances, similar to the picture of Fano-type resonances~\cite{fano1935,fano1961,joe2006}. The magnetic field and frequency values of the antiresonance are given by the following relations:
\begin{equation}
    \omega_{anti}=\frac{\gamma H_p}{1-\gamma m^* c/e}, \quad H_{anti}=\frac{H_p}{e/(\gamma m^* c)-1},
\end{equation}
which for the parameters of the system under consideration
give 6.84 GHz and 164 Oe, respectively.

To summarize, the absorption of an electromagnetic wave in a layered system consisting of a 2D electron system, a ferromagnetic film, and a metal placed in a perpendicular static magnetic field has been theoretically studied. The possibility of interaction (resulting in the appearance of an anticrossing) of the ferromagnetic resonance of the film and the cyclotron resonance of the electrons of the 2D system has been demonstrated. The parameters of the anticrossing (magnitude, magnetic field) as been determined. It has been found that the magnitude of the anticrossing for realistic parameters corresponding to Bi:TmIG films can reach several gigahertz.

However, it is worth clarifying that for practical observation of the interaction of the discussed resonances, the magnitude of the anticrossing must be larger than the broadening of the resonances, in particular, the collisional broadening of the cyclotron resonance, which is about $2/\tau$~\cite{Chiu:1976}, where $\tau$ is the electron relaxation time in the 2D system. Note that the time in record-mobility samples at low (liquid helium) temperatures can reach several nanoseconds \cite{chung2021}. The relaxation time in more typical samples is about 200 ps, which corresponds to a cyclotron resonance width about 1 GHz; i.e., it is of the same order as the value of the anticrossing in the structures proposed above. Therefore, to observe these effects at room temperature, we believe that antiferromagnetic films, the resonant frequencies of which are tens and hundreds of gigahertz \cite{Best_Review, Shao, Khymyn2017}, are more appropriate; since, the anticrossing between the cyclotron and magnetic resonances will be more pronounced. 

\begin{acknowledgments}
This work was carried out as part of the implementation of the Development Program of the World-Class Research Center “Center for Advanced Microelectronics” under agreement no. 075-15-2025-588 between the Ministry of Science and Higher Education of the Russian Federation and the Moscow Institute of Physics and Technology (National Research University).

\end{acknowledgments}

\end{document}